\documentclass[review]{elsarticle}

\usepackage{lineno,hyperref}
\modulolinenumbers[5]

\journal{Solid State Communications}

\begin{document}

\begin{frontmatter}

\title{Spin coherence as a function of depth for high-density ensembles of silicon vacancies in proton-irradiated 4H-SiC}

%

\author[mymainaddress]{P. G. Brereton\corref{mycorrespondingauthor}}
\cortext[mycorrespondingauthor]{Corresponding author}
\ead{brereton@usna.edu}

\author[mymainaddress]{D. Puent}

\author[mymainaddress]{J. Vanhoy}

\author[mysecondaryaddress]{E. R. Glaser}

\author[mysecondaryaddress]{S. G. Carter}

\address[mymainaddress]{Department of Physics, United States Naval Academy, Annapolis, Maryland, 21402, USA}
\address[mysecondaryaddress]{United States Naval Research Laboratory, 4555 Overlook Ave, Washington, DC, 20375, USA}

\begin{abstract}
Defects in wide-bandgap semiconductors provide a pathway for applications in quantum information and sensing in solid state materials.  The silicon vacancy in silicon carbide has recently emerged as a new candidate for optical control of single spin qubit with significant material benefits over nitrogen vacancies in diamond.  In this work, we present a study of the coherence of silicon vacancies generated via proton irradiation as a function of implantation depth.  We show clear evidence of dephasing interactions between the silicon vacancies and the spin environment of the bulk crystal.  This result will inform further routes toward fabrication of scalable silicon carbide devices and studies of spin interactions of in high-density ensembles of defects.
\end{abstract}


\end{frontmatter}


\section{Background}

Point defects in wide bandgap semiconductors are of great research interest in applications involving quantum information processing, quantum measurement and single photon emitters utilizing solid-state optoelectronic devices. The canonical solid state spin qubit has been the negatively charged nitrogen vacancy (NV) center in diamond.  Studies of NV centers have shown coherence times in excess of $100 \;\mu s$ \cite{Kennedy2003,Jelezko2004,Balasubramanian2009}, single photon generation \cite{Kurtsiefer2000}, nanoscale nuclear magnetic imaging \cite{Maletinsky2012,LeSage2013,Grinolds2013} and applications in quantum metrology \cite{Maze2008,Grinolds2014}. 
\subsection{SiC defects}

Diamond has significant drawbacks as a scalable platform for quantum technologies, however, including the high cost of fabrication and the relatively poor emission characteristics of the NV center zero-phonon line (ZPL).  Recently, significant focus has been brought to bear in a search for suitable defects in other wide-bandgap semiconductors \cite{Weber2010}. Several defects in silicon carbide (SiC) have emerged as candidates predicted to have spin characteristics similar to those found in diamond \cite{Koehl2011,Falk2013,Widmann2014a,Christle2014a,Lohrmann2015,Simin2016,Golter2017,Nagy2018}.  SiC offers significant advantages over diamond as a host for solid state spin qubits including a robust industrial capacity for high-throughput SiC microfabrication and emission in telecom compatible wavelength bands \cite{Caldwell2013,Calusine2014a}.  Additionally, SiC exists in over 250 polytypes, enabling the spin characteristics of each defect to be effectively “engineered” \cite{Falk2013}.

\subsection{High-density ensembles}

While significant effort has gone toward studies of the coherence of single SiC defects, little work has been done in the high-density ensemble limit. The radiative interactions of large ensembles of two-level optical systems have been the basis for ultra-stable lasers \cite{Bohnet2012} and quantum magnetometry \cite{Weiner2012}. Recently, superradiant coupling was demonstrated in the solid state for the first time in a dense ensemble of diamond NV centers \cite{Angerer2018}. In this report, we show that the $V_{Si}$ spin coherence time is strongly dependent on the depth within the sample and, thus the defect density. Furthermore, we show evidence that dipole-dipole interactions scale with defect density, thus indicating that proton irradiation is a viable route to generating high-density ensembles with strong radiative coupling.

\section{Methods and Materials}

In this report, we focus on the SiC silicon vacancy ($V_{Si}$) due to its low abundance in commercially available high purity SiC, emission in the near-infrared, and unique spin structure \cite{Sorman2000,Baranov2011,Riedel2012,Kraus2014,Carter2015a,Soykal2016}.  In the current published research on the silicon vacancy in SiC, four main defect generation methods have been employed: ion implantation \cite{Falk2015a,Falk2013}, electron beam irradiation \cite{Banks2019,Carter2015a}, neutron irradiation \cite{Kasper2019,Castelletto2013} and proton irradiation \cite{Ruehl2019,Ohshima2018,Embley2017,Kasper2019}.  For future device applications utilizing engineered defects in SiC, it is critical to identify the appropriate implementation protocols and the effect each may have on the spin coherence properties.  

\subsection{Proton irradiation}
The samples are commercially available (CREE) high purity semi-insulating 4H-SiC substrates.  Defects were generated by irradiation with a $2$ MeV proton beam from a tandem linear accelerator.  The samples were flood irradiated in air with the beam parallel to the c-axis of the bulk crystal.  Proton fluence in the sample used in this study was $3.5 \times 10^{15}$  proton/cm$^2$.  Ion stopping calculations yielded a projected proton range of $32 \; \mu$m into the bulk crystal \cite{Ziegler1985}.  The majority of the proton-induced defects of all types are expected to reside at this depth from the irradiated sample surface, with a smaller density of defects in the rest of the irradiated bulk. Unlike Ref. \cite{Ruehl2019}, no annealing step was performed and all emission is from natural and unannealed radiation damage defects alike.

\subsection{Photoluminecence}

\begin{figure}
    \includegraphics[width = \textwidth]{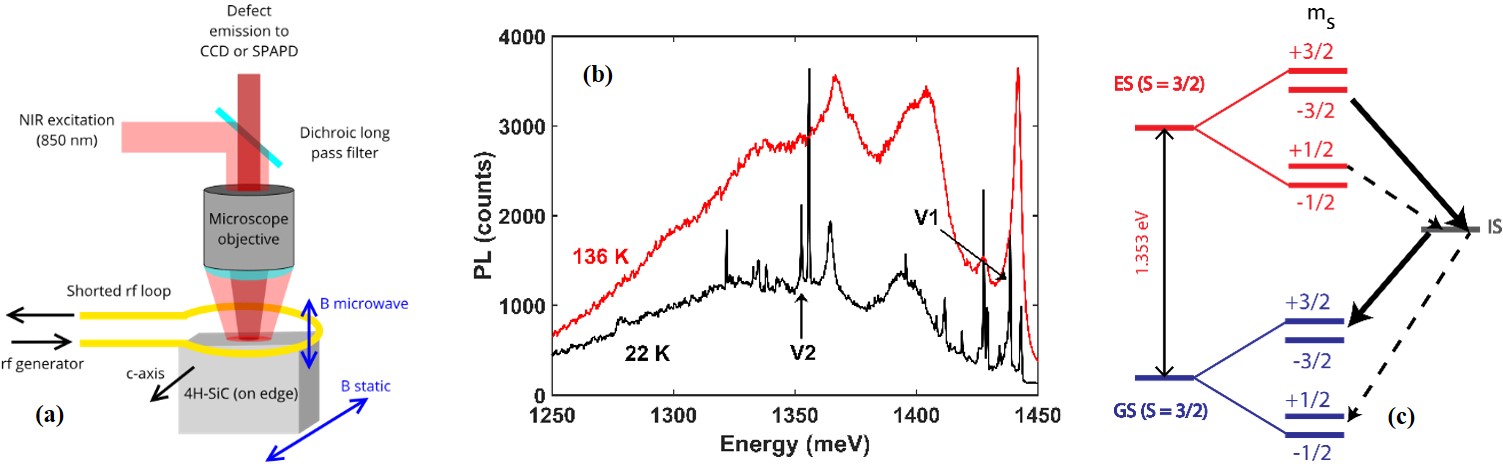}
    \caption{(a) Schematic of the experimental setup. (b) Typical PL spectra of irradiated SiC at 22K (black) and 136K (red) showing the ZPL of V1 and V2 and the broadening into phonon sidebands at elevated temperatures. (c) Energy level diagram of the V2 silicon vacancy. Optical transitions from the excited state (ES) to the ground state (GS) through the intersystem crossing (IS) are shown from each spin sub-state. The transitions shown by solid arrows occur at a faster rate than the dashed transitions.  See Ref. \cite{Soykal2016} for full details of the transition rates.}
    \label{fig:fig1}
\end{figure}

Defect photoluminescence (PL) is excited by below-gap excitation ($850$ nm) via a microscope objective ($0.65$ NA, $50 \times$), as illustrated in Fig. \ref{fig:fig1}(a).  Excitation light is generated from a tunable Ti:sapphire laser in either continuous wave or pulsed mode via a an acoustic optical modulator. The scattered defect emission is collected by the same objective, and sideband PL filtered by a long pass filter and sent either to a cooled, back-thinned Si CCD for spectroscopy or a Si avalanche photodiode for optically detected magnetic resonance (ODMR).  The excitation and collection are arranged confocally along the cleaved edge of the sample on a stepper motor stage to allow scanning from the irradiated face of the sample along its depth.  Microwave excitation is achieved by driving a loop of $50 \; \mu$m diameter gold wire shorted between the center electrode and the outer conductor of a coaxial cable with a radio frequency generator.  A static magnetic field is applied parallel to the c-axis via an external electromagnet.  Photoluminescence (PL) spectra at two different temperatures from a representative proton-irradiated 4H-SiC sample are shown below in Fig. \ref{fig:fig1}(b).

At low temperatures ($<25$ K), emission from mid-gap defects can be identified as narrow peaks in the near-IR.  The peaks at $1438$ meV and $1353$ meV (referred to as $V1$ and $V2$ respectively) are associated with the zero-phonon lines of transitions from the two non-equivalent silicon vacancy sites in the 4H polytype hexagonal lattice \cite{Kraus2014,Soykal2016}.  At room temperature (not shown), the emission in the phonon sidebands dominates and the zero-phonon lines broaden and weaken.  This work focuses on the $V2$ transition which has a strong spin-dependence in PL intensity at room temperature.

\section{Results}
\begin{figure}
    \includegraphics[width = \textwidth]{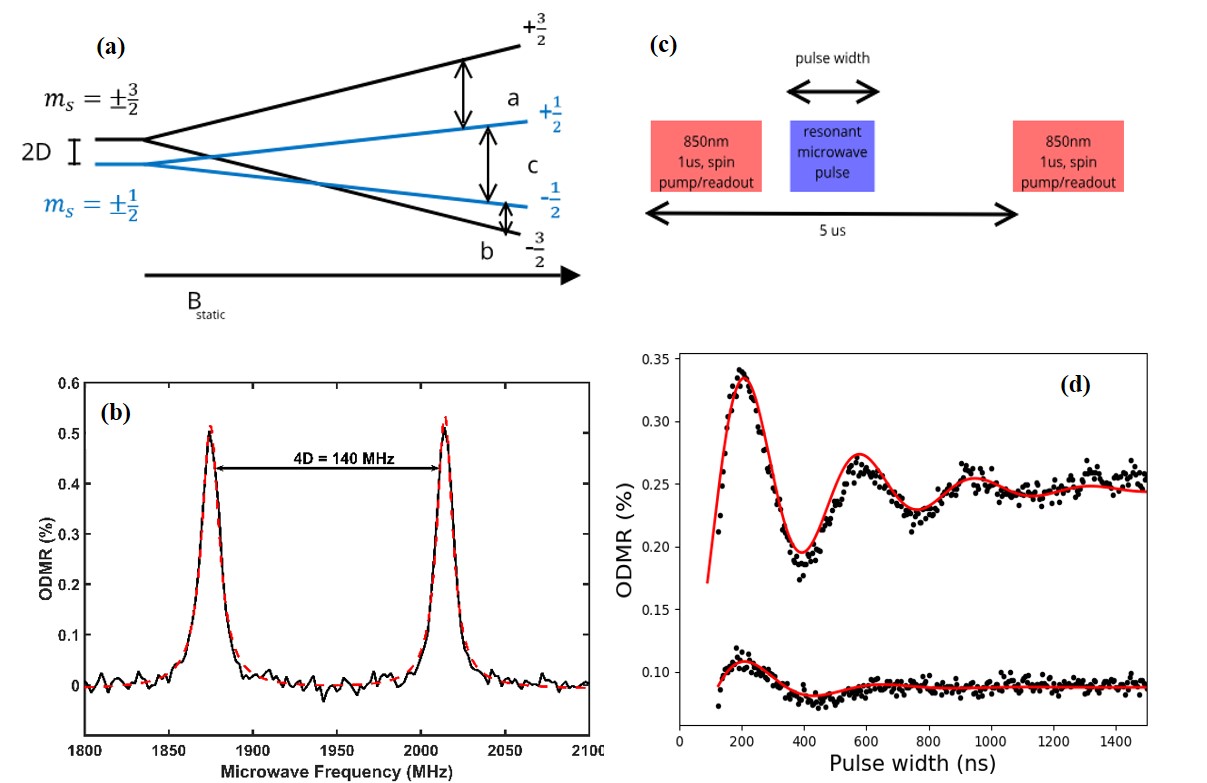}
    \caption{(a) Energy levels as a function of increasing static magnetic field.  Transitions a and b are ODMR-active, while c results in no change in PL intensity. (b) ODMR scan of a sample irradiated at $3.5 \times 10^{15}$  proton/cm$^2$ at a static field of $68.5$~mT. The red line is a Lorentzian fit. (c) Rabi pulse sequence. (d) Rabi rotations of the transition shown in 2(b) at $2022$ MHz at a depth of $13 \; \mu m$ (top) and $30\; \mu m$ (bottom). The red line is a fit to a damped sinusoid.}
    \label{fig:fig2}
\end{figure}

The spin state of the $V_{Si}$ is measured via the mechanism of optically detected magnetic resonance (ODMR).  As shown in the energy level diagram in Fig. \ref{fig:fig2}, the 4H silicon vacancy has a $S=  \frac{3}{2}$ ground state (GS) that is excited by optical excitation to a spin $\frac{3}{2}$ excited state (ES) \cite{Carter2015a,Soykal2017}.  Relaxation from the ES to the GS can occur through spin-preserving optical emission as well as a non-radiative pathway via intermediate states (IS) connecting different $m_s$ sub-states.

In the case of $V2$, the intensity of the PL is spin-dependent due to the higher probability of IS crossing transitions from  $m_s= \pm\frac{3}{2}$ to $\pm\frac{1}{2}$ \cite{Simin2016,Sorman2000}.  This also leads to a spin polarization of the $V_2$ defect via optical pumping of the GS preferentially into the $m_s= \pm\frac{1}{2}$ sublevel, a key requirement for optical spin control \cite{Baranov2011}. In the absence of an external magnetic field the splitting between the degenerate $m_s= \pm\frac{1}{2}$   and $\pm\frac{3}{2}$ sublevels in the GS is $2D=70$~MHz.  At non-zero fields parallel to the c-axis the spin sublevels split apart with a linear dependence on B, as shown in Fig. \ref{fig:fig2}(b) \cite{Soykal2016,Soykal2017}.  Dipole-allowed ($\Delta m_s= \pm1$) transitions between spin sublevels are labelled a, b and c in Fig. \ref{fig:fig2}(a).  In the ODMR mechanism, microwave radiation resonant with transition a or b will drive a transition from one spin sublevel to another, resulting in a change in PL intensity which is detected via a Si photodiode and a lock-in amplifier.  As shown in Fig. \ref{fig:fig2}(b), two transitions at $1883$ and $2022$~MHz are detected with a static magnetic field of $68.5$~mT under continuous wave optical excitation.   Transition c results in no change in spin state and thus no variation in PL intensity.

\subsection{Pulsed methods}

To assess the effect of defect density on the lifetime and coherence of defect spins, it is necessary to employ pulsed methods.  Rabi rotations are achieved by utilizing optical pulses to prepare the defect ensemble in an initial spin state and then vary the width of a microwave pulse resonant with transition a or b.  A final optical pulse measures the change in spin-dependent PL intensity.  As shown in Fig. \ref{fig:fig2}(d), the Rabi rotations resonant with the spin transition at $2022$~MHz under static magnetic field of $68.5$~mT are strongly damped due to the rapid inhomogeneous spin relaxation of the ensemble $T_2^*$ \cite{Carter2015a}.  This damping becomes significantly more pronounced as the maximum depth of proton implantation is approached, indicating an increase in inhomogeneous dephasing, potentially due to non-radiative interactions with other defects.

\begin{figure}
    \includegraphics[width = \textwidth]{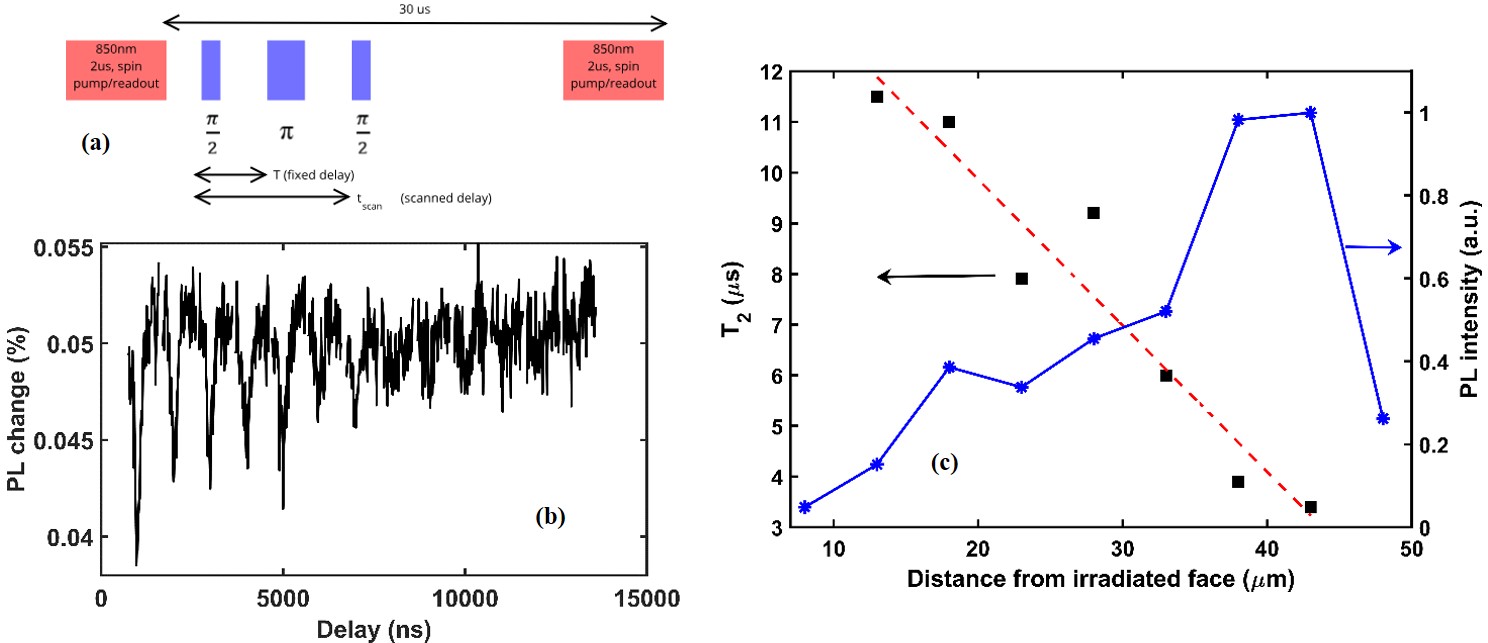}
    \caption{(a) Pulse sequence for the Hahn spin echo measurements. (b) Typical spin echo measurement at a depth of $8 \; \mu m$. (c) Coherence time ($T_2$) (left axis, the red dotted line is a guide to the eye) and total integrated defect PL intensity (right axis, blue solid line is a guide to the eye) as a function of depth within the sample.}
    \label{fig:fig3}
\end{figure}

\subsection{Coherence time as a function of depth}

In order remove the effect of inhomogeneous dephasing from the ensemble and measure the single defect coherence time ($T_2$), the Hahn echo separated pulse sequence is utilized \cite{Widmann2014a}.  In this method, as illustrated in Fig. \ref{fig:fig3}(a), an initial optical pump pulse prepares the ensemble into an initial spin polarization.  A $\frac{\pi}{2}$ resonant microwave pulse with a pulse width determined by the Rabi measurements rotates the ensemble spins into a superposition state.  After a fixed delay $T$ and subsequent free induction decay, a $\pi$ pulse reverses the polarity of the spins, followed by a final $\frac{\pi}{2}$ pulse after a variable delay time t.  The degree of coherence after the final microwave pulse is proportional to the change in the PL sideband intensity with a final off-resonant optical pulse.  As shown by the Hahn echo scan in Fig. \ref{fig:fig3}(b), the coherent echo signal at $t=2T$ damps out with increasing delays with characteristic coherence time $T_2$.  In this work, Hahn spin echo scans were performed at a series of depths within the bulk of the sample from just below the irradiated front surface to the calculated proton deposition depth.  As shown in Fig. \ref{fig:fig3}(c), we measured a near-monotonic decrease in $T_2$ from a maximum of $11.5 \; \mu s$ near the surface to $3.4 \; \mu s$ at a depth of approximately $40 \; \mu m$.  The integrated PL intensity at each depth increases as the stopping layer is approached, indicating an increase in the overall defect density that correlates with the decreasing coherence time.

\section{Conclusions}

\subsection{Dipole-dipole interaction}
Recent work on single $V_{Si}$ \cite{Widmann2014a} and defect ensembles \cite{Carter2015a} generated by electron irradiation as well as proton irradiation \cite{Embley2017} has indicated decoherence times on the order of $100 \; \mu s$ at room temperature. However, the coherence time has shown a strong dependence on defect density. For instance, Ref. \cite{Embley2017} and \cite{Kasper2019} show a decrease in $T_2$ from approximately $300 \; \mu s$ to $100 \; \mu s$ as proton fluence increases from $10^{13}$ to $10^{14}$ cm$^{-2}$. Thus, the surface $T_2 = 11.5 \; \mu s$ in our higher density sample fits this scaling law, indicating that dipole-dipole interactions with neighboring $V_{Si}$ is the dominant decoherence mechanism \cite{Embley2017}.  There is also a sharp decrease in $T_2$ as depth in the sample increases, indicating that the defect density may increase by an additional factor of $3$ in the vicinity of the stopping layer. This is critical in designing high-density ensembles of interacting dipoles. For instance, recent demonstrations of superradiance in diamond NV centers required an ensemble of $10^{16}$ interacting spins in a milimeter-sized mode volume \cite{Angerer2018}.

\subsection{Electron spin envelope modulation}
It has been shown that the envelope modulation of the spin echo shown in Fig. \ref{fig:fig3}(b) can be attributed to spin-spin interactions between the unpaired vacancy electrons and nearest neighbor nuclear spins \cite{Soykal2016,Soykal2017,Carter2015a}.  In our work, the delay time $T$ was short enough with respect to the dephasing time that this electron spin modulation could be neglected and a single exponential decay fit the spin echo. Furthermore, as shown in Ref. \cite{Carter2015a}, the spin echo revival at short delay times is suppressed by the application of a moderate external magnetic field, justifying our choice to ignore the envelope modification. However, as defect density increases toward the superradiant threshold, it will be important to quantify the impacts of nuclear hyperfine interactions on the ensemble coherence time. 

\subsection{Future Work}
Our work shows that proton irradiation is a viable route to generating high-density ensembles of $V_{Si}$ while providing a path toward engineering the coherence time of these defects through implantation depth.  Further work will be required to identify the specific dephasing mechanisms involved between defects and the proton-induced damage layer. Additionally, the relationship between irradiation energy (and thus depth) and defect density must be established. Finally, achieving superradiant emission from a dense ensemble of SiC defects is a critical step to realizing chip-scale ultra-stable lasers and detectors.

\vspace{1cm}

This research was supported by the Office of the Secretary of Defense Quantum Science and Engineering Program (QSEP). P.B. acknowledges the support received from the Office of Naval Research under grant N000141AF00002. D.P. was supported by the Bowman Scholarship at the United States Naval Academy.


\bibliography{mybibfile}

\begin{thebibliography}{10}
\expandafter\ifx\csname url\endcsname\relax
  \def\url#1{\texttt{#1}}\fi
\expandafter\ifx\csname urlprefix\endcsname\relax\def\urlprefix{URL }\fi
\expandafter\ifx\csname href\endcsname\relax
  \def\href#1#2{#2} \def\path#1{#1}\fi

\bibitem{Kennedy2003}
T.~A. Kennedy, J.~S. Colton, J.~E. Butler, R.~C. Linares, P.~J. Doering, Long
  coherence times at 300 k for nitrogen-vacancy center spins in diamond grown
  by chemical vapor deposition, Applied Physics Letters 83~(20) (2003)
  4190--4192.
\newblock \href {http://arxiv.org/abs/https://doi.org/10.1063/1.1626791}
  {\path{arXiv:https://doi.org/10.1063/1.1626791}}, \href
  {http://dx.doi.org/10.1063/1.1626791} {\path{doi:10.1063/1.1626791}}.

\bibitem{Jelezko2004}
F.~Jelezko, T.~Gaebel, I.~Popa, A.~Gruber, J.~Wrachtrup, Observation of
  coherent oscillations in a single electron spin, Phys. Rev. Lett. 92 (2004)
  076401.
\newblock \href {http://dx.doi.org/10.1103/PhysRevLett.92.076401}
  {\path{doi:10.1103/PhysRevLett.92.076401}}.

\bibitem{Balasubramanian2009}
G.~Balasubramanian, P.~Neumann, D.~Twitchen, M.~Markham, R.~Kolesov,
  N.~Mizuochi, J.~Isoya, J.~Achard, J.~Beck, J.~Tissler, et~al., Ultralong spin
  coherence time in isotopically engineered diamond, Nature Materials 8~(5)
  (2009) 383--387.
\newblock \href {http://dx.doi.org/10.1038/nmat2420}
  {\path{doi:10.1038/nmat2420}}.

\bibitem{Kurtsiefer2000}
C.~Kurtsiefer, S.~Mayer, P.~Zarda, H.~Weinfurter, Stable solid-state source of
  single photons, Phys. Rev. Lett. 85 (2000) 290--293.
\newblock \href {http://dx.doi.org/10.1103/PhysRevLett.85.290}
  {\path{doi:10.1103/PhysRevLett.85.290}}.

\bibitem{Maletinsky2012}
P.~Maletinsky, S.~Hong, M.~S. Grinolds, B.~Hausmann, M.~D. Lukin, R.~L.
  Walsworth, M.~Loncar, A.~Yacoby, A robust scanning diamond sensor for
  nanoscale imaging with single nitrogen-vacancy centres, Nature Nanotech 7~(5)
  (2012) 320--324.
\newblock \href {http://dx.doi.org/10.1038/nnano.2012.50}
  {\path{doi:10.1038/nnano.2012.50}}.

\bibitem{LeSage2013}
D.~Le~Sage, K.~Arai, D.~R. Glenn, S.~J. DeVience, L.~M. Pham, L.~Rahn-Lee,
  M.~D. Lukin, A.~Yacoby, A.~Komeili, R.~L. Walsworth, Optical magnetic imaging
  of living cells, Nature 496 (2013) 486.
\newblock \href {http://dx.doi.org/10.1038/nature12072}
  {\path{doi:10.1038/nature12072}}.

\bibitem{Grinolds2013}
M.~S. Grinolds, S.~Hong, P.~Maletinsky, L.~Luan, M.~D. Lukin, R.~L. Walsworth,
  A.~Yacoby, Nanoscale magnetic imaging of a single electron spin under ambient
  conditions, Nat Phys 9~(4) (2013) 215--219.
\newblock \href {http://dx.doi.org/10.1038/nphys2543}
  {\path{doi:10.1038/nphys2543}}.

\bibitem{Maze2008}
J.~R. Maze, P.~L. Stanwix, J.~S. Hodges, S.~Hong, J.~M. Taylor, P.~Cappellaro,
  L.~Jiang, M.~V.~G. Dutt, E.~Togan, A.~S. Zibrov, A.~Yacoby, R.~L. Walsworth,
  M.~D. Lukin, Nanoscale magnetic sensing with an individual electronic spin in
  diamond, Nature 455 (2008) 644.
\newblock \href {http://dx.doi.org/10.1038/nature07279}
  {\path{doi:10.1038/nature07279}}.

\bibitem{Grinolds2014}
M.~S. Grinolds, M.~Warner, K.~De~Greve, Y.~Dovzhenko, L.~Thiel, R.~L.
  Walsworth, S.~Hong, P.~Maletinsky, A.~Yacoby, Subnanometre resolution in
  three-dimensional magnetic resonance imaging of individual dark spins, Nature
  Nanotechnology 9~(4) (2014) 279???284.
\newblock \href {http://dx.doi.org/10.1038/nnano.2014.30}
  {\path{doi:10.1038/nnano.2014.30}}.

\bibitem{Weber2010}
J.~R. Weber, W.~F. Koehl, J.~B. Varley, A.~Janotti, B.~B. Buckley, C.~G. Van~de
  Walle, D.~D. Awschalom,
  \href{http://www.pnas.org/content/107/19/8513.abstract}{Quantum computing
  with defects}, Proceedings of the National Academy of Sciences 107~(19)
  (2010) 8513--8518.
\newblock \href
  {http://arxiv.org/abs/http://www.pnas.org/content/107/19/8513.full.pdf+html}
  {\path{arXiv:http://www.pnas.org/content/107/19/8513.full.pdf+html}}, \href
  {http://dx.doi.org/10.1073/pnas.1003052107}
  {\path{doi:10.1073/pnas.1003052107}}.
\newline\urlprefix\url{http://www.pnas.org/content/107/19/8513.abstract}

\bibitem{Koehl2011}
W.~F. Koehl, B.~B. Buckley, F.~J. Heremans, G.~Calusine, D.~D. Awschalom,
  \href{http://search.proquest.com/docview/905897309?accountid=14748}{Room
  temperature coherent control of defect spin qubits in silicon carbide},
  Nature 479~(7371) (2011) 84--7.
\newline\urlprefix\url{http://search.proquest.com/docview/905897309?accountid=14748}

\bibitem{Falk2013}
A.~L. Falk, B.~B. Buckley, G.~Calusine, W.~F. Koehl, V.~V. Dobrovitski,
  A.~Politi, C.~A. Zorman, P.~X.-L. Feng, D.~D. Awschalom, Polytype control of
  spin qubits in silicon carbide, Nature Communications 4 (2013) 1819.
\newblock \href {http://dx.doi.org/10.1038/ncomms2854}
  {\path{doi:10.1038/ncomms2854}}.

\bibitem{Widmann2014a}
M.~Widmann, S.-Y. Lee, T.~Rendler, N.~T. Son, H.~Fedder, S.~Paik, L.-P. Yang,
  N.~Zhao, S.~Yang, I.~Booker, et~al., Coherent control of single spins in
  silicon carbide at room temperature, Nature Materials\href
  {http://dx.doi.org/10.1038/nmat4145} {\path{doi:10.1038/nmat4145}}.

\bibitem{Christle2014a}
D.~J. Christle, A.~L. Falk, P.~Andrich, P.~V. Klimov, J.~U. Hassan, N.~Son,
  E.~Janz?©n, T.~Ohshima, D.~D. Awschalom, Isolated electron spins in silicon
  carbide with millisecond coherence times, Nature Materials\href
  {http://dx.doi.org/10.1038/nmat4144} {\path{doi:10.1038/nmat4144}}.

\bibitem{Lohrmann2015}
A.~Lohrmann, N.~Iwamoto, Z.~Bodrog, S.~Castelletto, T.~Ohshima, T.~J. Karle,
  A.~Gali, S.~Prawer, J.~C. McCallum, B.~C. Johnson, Single-photon emitting
  diode in silicon carbide, Nature Communications 6 (2015) 7783.
\newblock \href {http://dx.doi.org/10.1038/ncomms8783}
  {\path{doi:10.1038/ncomms8783}}.

\bibitem{Simin2016}
D.~Simin, V.~A. Soltamov, A.~V. Poshakinskiy, A.~N. Anisimov, R.~A. Babunts,
  D.~O. Tolmachev, E.~N. Mokhov, M.~Trupke, S.~A. Tarasenko, A.~Sperlich, P.~G.
  Baranov, V.~Dyakonov, G.~V. Astakhov, All-optical dc nanotesla magnetometry
  using silicon vacancy fine structure in isotopically purified silicon
  carbide, Phys. Rev. X 6 (2016) 031014.
\newblock \href {http://dx.doi.org/10.1103/PhysRevX.6.031014}
  {\path{doi:10.1103/PhysRevX.6.031014}}.

\bibitem{Golter2017}
D.~A. Golter, C.~W. Lai, Optical switching of defect charge states in 4h-sic,
  Scientific Reports 7~(1) (2017) 13406.
\newblock \href {http://dx.doi.org/10.1038/s41598-017-13813-2}
  {\path{doi:10.1038/s41598-017-13813-2}}.

\bibitem{Nagy2018}
R.~Nagy, M.~Widmann, M.~Niethammer, D.~B.~R. Dasari, I.~Gerhardt, O.~O. Soykal,
  M.~Radulaski, T.~Ohshima, J.~Vu\ifmmode \check{c}\else
  \v{c}\fi{}kovi\ifmmode~\acute{c}\else \'{c}\fi{}, N.~T. Son, I.~G. Ivanov,
  S.~E. Economou, C.~Bonato, S.-Y. Lee, J.~Wrachtrup, Quantum properties of
  dichroic silicon vacancies in silicon carbide, Phys. Rev. Applied 9 (2018)
  034022.
\newblock \href {http://dx.doi.org/10.1103/PhysRevApplied.9.034022}
  {\path{doi:10.1103/PhysRevApplied.9.034022}}.

\bibitem{Caldwell2013}
J.~D. Caldwell, O.~J. Glembocki, Y.~Francescato, N.~Sharac, V.~Giannini, F.~J.
  Bezares, J.~P. Long, J.~C. Owrutsky, I.~Vurgaftman, J.~G. Tischler, V.~D.
  Wheeler, N.~D. Bassim, L.~M. Shirey, R.~Kasica, S.~A. Maier, Low-loss,
  extreme subdiffraction photon confinement via silicon carbide localized
  surface phonon polariton resonators, Nano Letters 13~(8) (2013) 3690--3697,
  pMID: 23815389.
\newblock \href {http://arxiv.org/abs/https://doi.org/10.1021/nl401590g}
  {\path{arXiv:https://doi.org/10.1021/nl401590g}}, \href
  {http://dx.doi.org/10.1021/nl401590g} {\path{doi:10.1021/nl401590g}}.

\bibitem{Calusine2014a}
G.~Calusine, A.~Politi, D.~D. Awschalom, Silicon carbide photonic crystal
  cavities with integrated color centers, Applied Physics Letters 105~(1)
  (2014) 011123.
\newblock \href {http://dx.doi.org/10.1063/1.4890083}
  {\path{doi:10.1063/1.4890083}}.

\bibitem{Bohnet2012}
J.~G. Bohnet, Z.~Chen, J.~M. Weiner, D.~Meiser, M.~J. Holland, J.~K. Thompson,
  A steady-state superradiant laser with less than one intracavity photon,
  Nature 484~(7392) (2012) 78--81.
\newblock \href {http://dx.doi.org/10.1038/nature10920}
  {\path{doi:10.1038/nature10920}}.

\bibitem{Weiner2012}
J.~M. Weiner, K.~C. Cox, J.~G. Bohnet, Z.~Chen, J.~K. Thompson, Superradiant
  raman laser magnetometer, Applied Physics Letters 101~(26) (2012) 261107.
\newblock \href {http://arxiv.org/abs/https://doi.org/10.1063/1.4773241}
  {\path{arXiv:https://doi.org/10.1063/1.4773241}}, \href
  {http://dx.doi.org/10.1063/1.4773241} {\path{doi:10.1063/1.4773241}}.

\bibitem{Angerer2018}
A.~Angerer, K.~Streltsov, T.~Astner, S.~Putz, H.~Sumiya, S.~Onoda, J.~Isoya,
  W.~J. Munro, K.~Nemoto, J.~Schmiedmayer, J.~Majer, Superradiant emission from
  colour centres in diamond, Nature Physics 14~(12) (2018) 1168--1172.
\newblock \href {http://dx.doi.org/10.1038/s41567-018-0269-7}
  {\path{doi:10.1038/s41567-018-0269-7}}.

\bibitem{Sorman2000}
E.~S\"orman, N.~T. Son, W.~M. Chen, O.~Kordina, C.~Hallin, E.~Janz\'en, Silicon
  vacancy related defect in 4h and 6h sic, Phys. Rev. B 61 (2000) 2613--2620.
\newblock \href {http://dx.doi.org/10.1103/PhysRevB.61.2613}
  {\path{doi:10.1103/PhysRevB.61.2613}}.

\bibitem{Baranov2011}
P.~G. Baranov, A.~P. Bundakova, A.~A. Soltamova, S.~B. Orlinskii, I.~V.
  Borovykh, R.~Zondervan, R.~Verberk, J.~Schmidt, Silicon vacancy in sic as a
  promising quantum system for single-defect and single-photon spectroscopy,
  Phys. Rev. B 83 (2011) 125203.
\newblock \href {http://dx.doi.org/10.1103/PhysRevB.83.125203}
  {\path{doi:10.1103/PhysRevB.83.125203}}.

\bibitem{Riedel2012}
D.~Riedel, F.~Fuchs, H.~Kraus, S.~V\"ath, A.~Sperlich, V.~Dyakonov, A.~A.
  Soltamova, P.~G. Baranov, V.~A. Ilyin, G.~V. Astakhov, Resonant addressing
  and manipulation of silicon vacancy qubits in silicon carbide, Phys. Rev.
  Lett. 109 (2012) 226402.
\newblock \href {http://dx.doi.org/10.1103/PhysRevLett.109.226402}
  {\path{doi:10.1103/PhysRevLett.109.226402}}.

\bibitem{Kraus2014}
H.~{Kraus}, V.~A. {Soltamov}, D.~{Riedel}, S.~{V{\"a}th}, F.~{Fuchs},
  A.~{Sperlich}, P.~G. {Baranov}, V.~{Dyakonov}, G.~V. {Astakhov},
  {Room-temperature quantum microwave emitters based on spin defects in silicon
  carbide}, Nature Physics 10 (2014) 157--162.
\newblock \href {http://dx.doi.org/10.1038/nphys2826}
  {\path{doi:10.1038/nphys2826}}.

\bibitem{Carter2015a}
S.~G. Carter, O.~O. Soykal, P.~Dev, S.~E. Economou, E.~R. Glaser, Spin
  coherence and echo modulation of the silicon vacancy in $4h-\mathrm{SiC}$ at
  room temperature, Phys. Rev. B 92 (2015) 161202.
\newblock \href {http://dx.doi.org/10.1103/PhysRevB.92.161202}
  {\path{doi:10.1103/PhysRevB.92.161202}}.

\bibitem{Soykal2016}
O.~O. Soykal, P.~Dev, S.~E. Economou, Silicon vacancy center in $4h$-sic:
  Electronic structure and spin-photon interfaces, Phys. Rev. B 93 (2016)
  081207.
\newblock \href {http://dx.doi.org/10.1103/PhysRevB.93.081207}
  {\path{doi:10.1103/PhysRevB.93.081207}}.

\bibitem{Falk2015a}
A.~L. Falk, P.~V. Klimov, V.~Iv\'ady, K.~Sz\'asz, D.~J. Christle, W.~F. Koehl,
  A.~Gali, D.~D. Awschalom, Optical polarization of nuclear spins in silicon
  carbide, Phys. Rev. Lett. 114 (2015) 247603.
\newblock \href {http://dx.doi.org/10.1103/PhysRevLett.114.247603}
  {\path{doi:10.1103/PhysRevLett.114.247603}}.

\bibitem{Banks2019}
H.~B. Banks, O.~O. Soykal, R.~L. Myers-Ward, D.~K. Gaskill, T.~Reinecke, S.~G.
  Carter, Resonant optical spin initialization and readout of single silicon
  vacancies in $4h$-$\mathrm{Si}\mathrm{C}$, Phys. Rev. Applied 11 (2019)
  024013.
\newblock \href {http://dx.doi.org/10.1103/PhysRevApplied.11.024013}
  {\path{doi:10.1103/PhysRevApplied.11.024013}}.

\bibitem{Kasper2019}
C.~Kasper, D.~Klenkert, Z.~Shang, D.~Simin, A.~Sperlich, H.~Kraus,
  C.~Schneider, S.~Zhou, M.~Trupke, W.~Kada, T.~Ohshima, V.~Dyakonov, G.~V.
  Astakhov, Influence of irradiation on defect spin coherence in silicon
  carbide\href {http://arxiv.org/abs/http://arxiv.org/abs/1908.06829v1}
  {\path{arXiv:http://arxiv.org/abs/1908.06829v1}}.

\bibitem{Castelletto2013}
S.~Castelletto, B.~C. Johnson, V.~Iv??dy, N.~Stavrias, T.~Umeda, A.~Gali,
  T.~Ohshima, A silicon carbide room-temperature single-photon source, Nature
  Materials 13~(2) (2013) 151???156.
\newblock \href {http://dx.doi.org/10.1038/nmat3806}
  {\path{doi:10.1038/nmat3806}}.

\bibitem{Ruehl2019}
M.~Rühl, C.~Ott, S.~Götzinger, M.~Krieger, H.~B. Weber,
  \href{https://doi.org/10.1063/1.5045859}{Controlled generation of intrinsic
  near-infrared color centers in 4h-sic via proton irradiation and annealing},
  Appl. Phys. Lett. 113~(12) (2019) 122102.
\newblock \href {http://dx.doi.org/10.1063/1.5045859}
  {\path{doi:10.1063/1.5045859}}.
\newline\urlprefix\url{https://doi.org/10.1063/1.5045859}

\bibitem{Ohshima2018}
T.~Ohshima, T.~Satoh, H.~Kraus, G.~V. Astakhov, V.~Dyakonov, P.~G. Baranov,
  \href{https://doi.org/10.1088%2F1361-6463%2Faad0ec}{Creation of silicon
  vacancy in silicon carbide by proton beam writing toward quantum sensing
  applications}, Journal of Physics D: Applied Physics 51~(33) (2018) 333002.
\newblock \href {http://dx.doi.org/10.1088/1361-6463/aad0ec}
  {\path{doi:10.1088/1361-6463/aad0ec}}.
\newline\urlprefix\url{https://doi.org/10.1088%2F1361-6463%2Faad0ec}

\bibitem{Embley2017}
J.~S. Embley, J.~S. Colton, K.~G. Miller, M.~A. Morris, M.~Meehan, S.~L.
  Crossen, B.~D. Weaver, E.~R. Glaser, S.~G. Carter, Electron spin coherence of
  silicon vacancies in proton-irradiated 4h-sic, Phys. Rev. B 95 (2017) 045206.
\newblock \href {http://dx.doi.org/10.1103/physrevb.95.045206}
  {\path{doi:10.1103/physrevb.95.045206}}.

\bibitem{Ziegler1985}
J.~F. Ziegler, J.~P. Biersack,
  \href{https://doi.org/10.1007/978-1-4615-8103-1_3}{The stopping and range of
  ions in matter}, in: D.~A. Bromley (Ed.), Treatise on Heavy-Ion Science:
  Volume 6: Astrophysics, Chemistry, and Condensed Matter, Springer US, Boston,
  MA, 1985, pp. 93--129.
\newline\urlprefix\url{https://doi.org/10.1007/978-1-4615-8103-1_3}

\bibitem{Soykal2017}
O.~O. Soykal, T.~L. Reinecke,
  \href{https://link.aps.org/doi/10.1103/PhysRevB.95.081405}{Quantum metrology
  with a single spin-$\frac{3}{2}$ defect in silicon carbide}, PRB 95~(8)
  (2017) 081405.
\newline\urlprefix\url{https://link.aps.org/doi/10.1103/PhysRevB.95.081405}

\end{thebibliography}

\end{document}